\documentclass[prd,nofootinbib,showpacs,12pt]{revtex4}

\usepackage{graphicx}

\begin{document}


\title{Critical Phenomena in DIS}

\author{L.L.~Jenkovszky\,$^a$}\email{jenk@bitp.kiev.ua}
\author{Andrea Nagy\,$^b$}\email{tarics1@rambler.ru}
\author{S.M.~Troshin\,$^c$}\email{sergey.troshin@ihep.ru}
\author{Jol\'{a}n Tur\'{o}ci\,$^b$}\email{tarics1@rambler.ru}
\author{N.E.~Tyurin\,$^c$}\email{tyurin@ihep.ru}
\affiliation{
$^a$BITP, Kiev, 03680 Ukraine\\
$^b$Uzhgorod State University, Ukraine\\
$^c$IHEP, Protvino, 142284 Russia}
\begin{abstract}
Saturation in deep inelastic scattering (DIS) and deeply virtual
Compton scattering (DVCS) is associated with a phase transition
between the partonic gas, typical of moderate  $x$ and $Q^2$, and
partonic fluid appearing at increasing $Q^2$ and decreasing
Bjorken $x$. In this paper we do not intend to propose another
parametrization of the structure function; instead we suggest a
new insight into the internal structure of the nucleon, as seen in
DIS, and its connection with that revealed in high-energy nucleons
and heavy-ion collisions.
\end{abstract}


\maketitle
\section{Introduction} \label{I}
The thermodynamic approach to high-energy hadronic collisions was
initiated by the paper of Fermi Ref. \cite{fermi}, subsequently
successfully applied to high-energy nuclear collisions. Less
familiar, although numerous, are the applications of the
thermodynamical approach to deep inelastic scattering (DIS)
\cite{Bhalerao, Cleymans, Soffer, JTT, Dubna}. Formally, by vector
meson dominance, DIS can also fall in the category of
hadron-hadron (heavy vector meson-baryon) processes. While the
creation of a large number of particles, justifying the
applicability of the statistical mechanics, is typical of both
classes of reactions, there are subtleties that make the
thermodynamics of DIS different from that of hadronic or nuclear
collisions. One is connected with the reference frame: while
thermalization and possible creation of "quark-gluon plasma" is
considered in the rest frame, the partonic picture of DIS is
realized in the infinite momentum frame, $p_z\rightarrow\infty.$
The connection and transition between these frames recently was
treated in Ref. \cite{Dubna}. The second point is the role and
treatment of the temperature. Any statistical distribution in a
gas of quasi-free particles (nucleons or partons) implies the
existence of a "temperature" $T$, as e.g. in the Boltzmann
distribution $e^{-E/T}$, although its physical interpretation is
not unique. A limited temperature $T$ is typical of the hadronic
phase \cite{Hagedorn} but not of the partons, that can be heated
indefinitely.

In Ref. \cite{Krzywicki} arguments based on parton-hadron duality
lead the concept of "pseudo-thermalization", according to which,
after the hadron is broken by the interaction, the inclusive
spectrum reflects the "thermal" distribution holding at the parton
level.

A related problem in DIS is the choice of the variable in the
statistical distribution. The use of the Bjorken variable $x,$
instead of the energy (or momentum), in the above (or similar)
distributions needs the introduction of a proper dimensional
parameter, related to the change of the coordinate systems (see,
e.g. \cite{Dubna}), $\exp(-x/\bar x),\ \ \bar x=x(1+k^2/x^2m^2),$
where $k$ is the quark transverse momentum and $m$ is the proton
mass. Note that the "temperature" $T$ here is that of the partonic
gas (or liquid/fluid) inside the nucleon and hence, unlike the
hadronic systems \cite{Hagedorn}, it must not be limited.

\begin{figure}[h]
\begin{center}
\hspace{-1.cm}
\includegraphics[width=1\textwidth,angle=0]{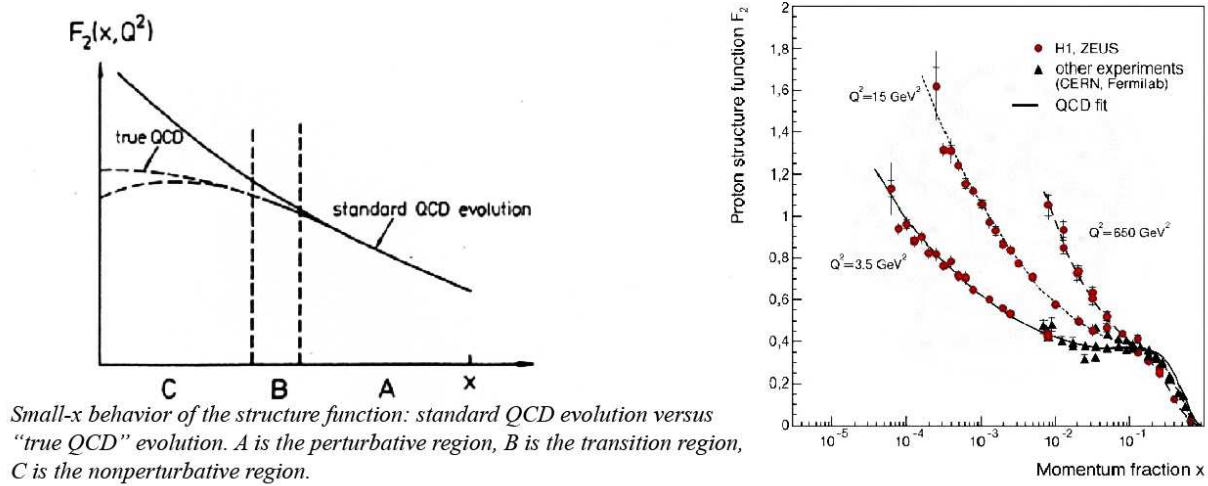}
\caption{\small \it {Change of regime in the behavior of the SF is
visible around $x\sim 10^{-2}$. Beyond that value, Bjorken scaling
holds (modulus $\ln Q^2$), while at lower $x$, the $Q^2$
dependence changes drastically and it cannot be considered as a
small scaling violating effect anymore (see Ref.
\cite{Yoshida}.)}} \label{fig:Yoshida}
\end{center}
\end{figure}

It was suggested in Ref. \cite{JTT} that saturation in DIS,
predicted by QCD and observed experimentally, corresponds to the
condensation of partonic gas to a fluid. The interior of the
nucleon excited in DIS or DVCS undergoes a phase transition from a
partonic gas (high- and intermediate, $x\ \ x\sim 0.05$) to a
partonic fluid. The division line is located roughly at those
values of $x$ and $Q^2$ where Bjorken scaling is not valid any
more, as shown in Fig. \ref{fig:Yoshida}.

Our idea \cite{JTT}, based on the observed behavior of the DIS
structure functions, is that the partonic matter in a nucleon (or
nucleus) as seen in DIS, undergoes a change of phase from a nearly
perfect gas, typical of the Bjorken scaling region, to a liquid,
where the logarithmic scaling violation is replaced by a power
(see Figure \ref{fig:Yoshida}). The presence of two regions,
namely that of Bjorken scaling and beyond it (call them for the
moment "dilute" and "dense"), via an intermediate mixed phase, are
visible in this figure. They can be quantifies in various ways
that will be discussed in the next section. The relevant variable
here are the fraction of the nucleon momentum, or the Bjorken
variable $x$ and the incident photon's virtuality $Q^2$.

\begin{figure}[h]
\begin{center}
\includegraphics[width=0.8\textwidth,angle=0]{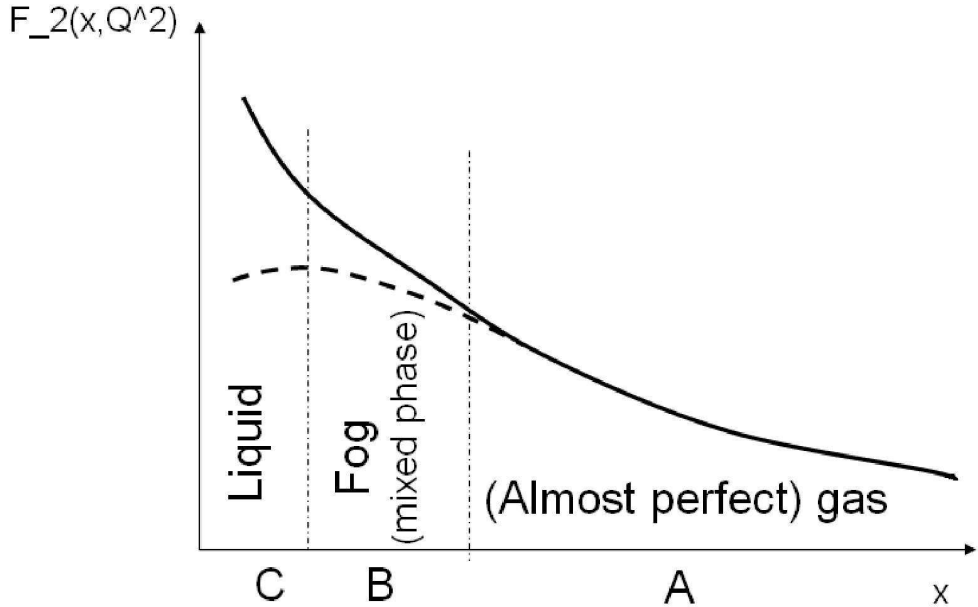}
\caption{\small \it {Gaseous, foggy and liquid states in various
kinematical regions of DIS (cf. Fig. \ref{fig:Yoshida}).}}
\label{SF1}
\end{center}
\end{figure}

The co-existence of two phases, gaseous and fluid, can be
described e.g. by the van der Waals equation, valid in a
tremendous range of its variables and applicable to any system
(e.g. molecular, atomic or nuclear) with short range repulsion and
long range attraction between the constituents, see e.g. Fig.
\ref{SF1} from Ref. \cite{Jaqaman}.

\begin{figure}[h]
\begin{center}
\hspace{-1.cm}
\includegraphics[width=0.9\textwidth,angle=0]{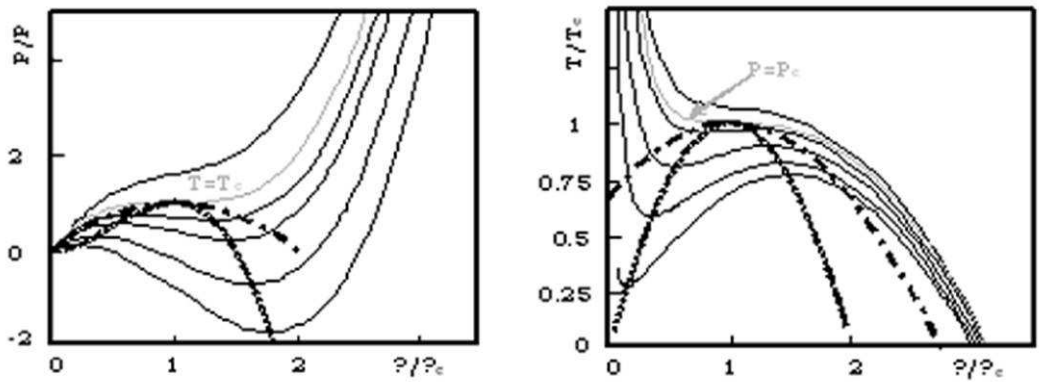}
\caption{\small \it {Universal van der Waals curves from Ref.
\cite{Jaqaman}.}} \label{Jaqaman}
\end{center}
\end{figure}

In Sec. \ref{IV} we describe the properties of the partonic
matter by means of the van der Waals equation and discuss the
physical meaning of its parameters. To anticipate, the equation of
state shows two phases, gaseous and fluid, with a mixed phase
in-between, called spinodal.

Notice that the two phases, gaseous and fluid, here are made of
the same stuff - colored quarks and gluons. We do not consider the
possibility of a (de)confinement phase transition. In this sense,
the situation is similar to that "low-energy" nuclear physics or
in classical substances, such as vapour condensation.

Comparison between the two pictures, the DIS structure functions
and the van der Waals EOS, especially at the critical points
(regions), is the subject of our discussion in Secs. \ref {IV}
and \ref{Mapping} .

\section{DIS Structure functions, geometrical scaling and saturation}\label{II}
Most generally, DIS structure functions (SF) are sums of a singlet
(S) and non-singlet (NS) terms,
$F_2(x,Q^2)=F_2^S(x,Q^2)+F_2^{NS}(x,Q^2),$ each a product of a
low-$x,\sim x^\alpha$ and high-$x,\ (1-x)^n$ factor, more
specifically (see e.g. \cite{Rivista} for more details):

\begin{equation}\label{SF} F_2^S(x,Q^2)=
A_0\Biggl({Q^2\over{Q^2+a}}\Biggr)^{1+\Delta(Q^2)}x^{-\Delta(Q^2)}(1-x)^{n(Q^2)+4},\
\
\Delta(Q^2)=\Delta_0\Biggl(1+{bQ^2\over{Q^2+c}}\Biggr),\end{equation}
where $\Delta_0\approx 0.1,\ \ b\approx 0.4,\ etc.$ The above
model is applicable in the Regge domain of small and intermediate
values of $Q^2$, mimicking the apparent "hardening" of the
Pomeron, manifest in the rise of $\Delta(Q^2)$ from about $0.1$ to
$0.4$. At higher virtualities, Regge behavior should be replaced
by the effects of QCD evolution, absent from the above simple
model. The drastic increase of $F_2(x,Q^2)$ in $1/x$ with
increasing $Q^2$, as seen in Fig. \ref{fig:Yoshida}, indicates the
onset of a new dynamical regime. Although, for virtual particles,
i.e. in DIS, there is no analogue of the Froissart bound, there
are arguments why a change of regime should occur.

\begin{figure}[h]
\begin{center}
\hspace{-1.cm}
\includegraphics[width=0.64\textwidth,angle=0]{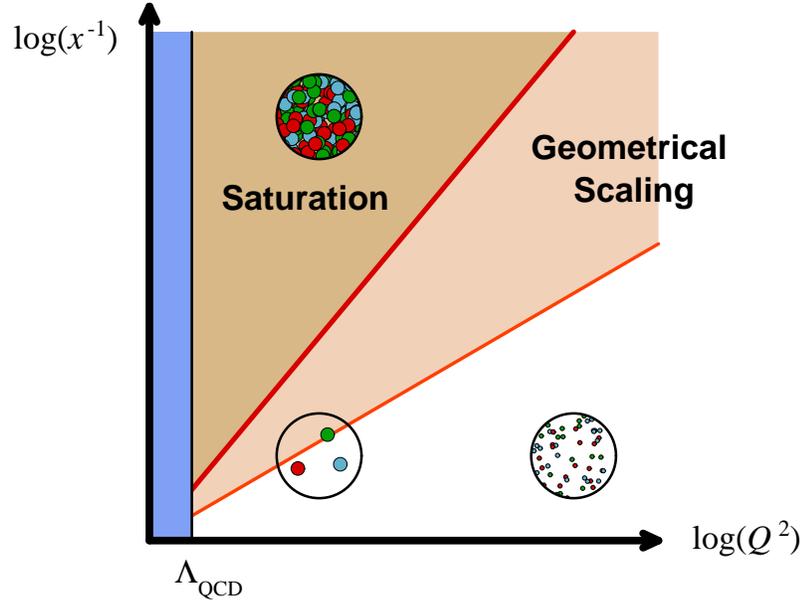}
\caption{\small \it {Phase diagram of DIS, from Ref.
\cite{Gelis}.}} \label{fig:satur}
\end{center}
\end{figure}

A simple and convincing argument is a physical one, according to
which, by the drastically (power-like) scaling-violating rise of
$F_2(x,Q^2)$, the number of partons is increasing and their volume
tends to exceed that of the nucleon (gluon saturation), thus
leading to partons' recombination (or their condensation, in terms
of statistical physics, see below). To quantify this effect, one
compares the number of gluons per unit of transverse
area,\footnote{Due to Lorentz contraction in the longitudinal
direction, soft gluons belonging to different nucleons have
overlapping wave functions and thus act coherently. The effect of
transverse (impact-parameter) distribution can be deduced from
DVCS scattering and general parton distributions \cite{TT, Jenk}}
$\rho\sim xG(x,Q^2)/\pi R^2,$ and the cross section for
recombination, $\sigma\sim\alpha_s/Q^2,$ where $\alpha_s$ is the
QCD running coupling. Saturation occurs when
$1\lesssim\rho\sigma,$ or equivalently $Q^2\leq Q^2_s(x),$ where
$Q^2_s(x)= \alpha_sxG(xQ^2)/\pi R^2$ is the so-called saturation
momentum \cite{Golec}. The saturation domain, shown in Fig.
\ref{fig:satur}, is delineated by the equation $Q^2=Q^2_s(x)$;
phenomenologically \cite{Golec}, $Q^2_s(x)\sim x^{-0.3}$.

A related phenomenon in DIS is the so-called geometrical scaling
(GS) (not to be confused with GS in hadronic processes!), implying
the existence of a "saturation radius" $R_0(x),$ given by
\cite{Golec}
\begin{equation}\label{GS}
R_0^2(x)=\Bigl({{x}\over{x_0}}\Bigr)^\lambda/Q^2_0,
\end{equation} where $Q^2_0=1$ GeV$^2,\ \ \lambda=0.29$
and $x_0=3\cdot10^{-4}$. GS was derived \cite{Golec} from the
dipole model of DIS and was shown to be compatible with the HERA
data in a wide span of $x$ and $Q^2$.

Substitution of (\ref{GS}) into the effective intercept of the
low-$x$ factor in Eq. (\ref{SF}) (the Pomeron contribution),
\begin{equation}\label{satur}
F_2(x,Q^2)\sim x^{\Delta[Q^2_s(x)\sim x^{-0.3}]},
\end{equation}
produces a maximum (saturation) in
this simple model for the singlet (gluonic) structure function.

Formally, saturation can be treated in the context of the BFKL
equation \cite{BFKL} and its modifications~\cite{Kovcheg}. The
production of partons in a nucleon by splitting of partons,
resulting in the increase of their number $N,$ is described by the
asymptotic BFKL evolution equation
\begin{equation}\label{BFKL}
\frac{\partial
N(x,k_T^2)}{\partial\ln(1/x)}=\alpha_sK_{BFKL}\otimes
N(x,K_T^2),
\end{equation}
where $K_{BFKL}$ is the BFKL integral kernel (splitting function).
It results in a power-like increase of the SF toward smaller $x,\
\ F_2(x)\sim x^{-\alpha_P+1},$ where $\alpha_P$ is the BFKL
pomeron intercept, $\alpha_P-1=(4\alpha_sN_c\ln 2)/\pi>0;\ \
\alpha_s$ is the ("running") QCD coupling and $N_c$ is the number
of colours. Since the number of partons increases with energy, at
certain "saturation" values of $x$ and $Q^2,$ an inverse process,
namely the recombination of pairs of partons comes into play, and
the BFKL equation is replaced by the following one \cite{BK}
\begin{equation}\label{BK}
\frac{\partial
N(x,k_T^2)}{\partial\ln(1/x)}=\alpha_sK_{BFKL}\otimes
N(x,K_T^2)-\alpha_s[K_{BFKL}\otimes N(x,K_T^2)]^2,
\end{equation}
based on the simple idea that the number of recombinations is
roughly proportional to the the number of parton pairs, $N^2$.
Saturation sets in when the second, quadratic term in Eq.
(\ref{BK}) overshoots the first, linear one, thus tampering the
increase of the number of produced partons and securing unitarity.

An explicit model realizing the onset of saturation, based on an
interpolation between the known asymptotic -- Regge and DGLAP -
regimes of the structure functions, yet fitting the data, was
proposed in Ref. \cite{Interpolate}. This model will be used to
match the saturation region in DIS with the predicted phase
transition.

The ansatz for the small-$x$ singlet part (labelled by the upper
index ($S,0$) of the proton structure function, interpolating
between the soft (VMD, Pomeron) and hard (DGLAP evolution)
regimes, reads \cite{Interpolate}:

\begin{equation}\label{Inter}F_{2}^{(S,0)}(x,Q^2) =
A\left({Q^2\over Q^2+a}\right)^{1+\widetilde{\Delta} (Q^2)}
e^{\Delta (x,Q^2)},\end{equation} with the "effective power"
$$ \widetilde{\Delta }(Q^2) =\epsilon+\gamma_1\ell n {
\left(1+\gamma_2\ell n{\left[1+{Q^2\over Q^2_0}\right]}\right)},$$
and
$$\Delta (x,Q^2) = \left(\widetilde{\Delta } (Q^2)  \ell n{x_0\over x}\right)^{f(Q^2)},$$
where
$$ f(Q^2) = {1\over 2}\left( {1+e^{-{Q^2/Q_1^2}}}\right). $$

At small and moderate values of $Q^2$ (specified
\cite{Interpolate} by fits to the data), the exponent
$\widetilde{\Delta}(Q^2)$  may be interpreted as a $Q^2$-dependent
"effective Pomeron intercept".

The function $f(Q^2)$ "swithches" Regge behavior, where
$f(Q^2)=1$, to the asymptotic solution of the GLAP evolution
equation, where $f(Q^2)=1/2$.

By construction, the model has the following asymptotic limits:

a) Large $Q^2$, fixed $x$:
$$ F_{2}^{(S,0)}(x,Q^2\to \infty)\to A\
\exp^{\sqrt{\gamma_1\ell n\ell n{Q^2\over Q_0^2}\ \ell n{x_0\over
x}}}\ ,$$ which is the asymptotic solution of the GLAP evolution
equation.

b) Low $Q^2$, fixed $x$:
$$F_{2}^{(S,0)}(x,Q^2\to 0) \to A\
e^{\Delta (x,Q^2\to 0)} \ \left({Q^2\over
a}\right)^{1+\widetilde{\Delta }(Q^2\to 0)}$$ with
$$ \widetilde{\Delta }(Q^2\to 0) \to \epsilon+\gamma_1 \gamma_2
{ \left({{Q^2\over Q^2_0} }\right)}\ \to\ \epsilon ,$$
$$ f(Q^2\to 0) \to 1 ,$$
where from
$$F_{2}^{(S,0)}(x,Q^2\to 0) \to A\ \left( {x_0\over x}
\right)^\epsilon \ \left({Q^2\over a}\right)^{1+\epsilon} \
\propto (Q^2)^{1+\epsilon} \ \to 0\ ,$$ as required by gauge
invariance.

c) Low $x$, fixed $Q^2$:
$$F_{2}^{(S,0)}(x\to 0,Q^2) \ =\
A\left({Q^2\over Q^2 + a}\right)^{1+\widetilde{\Delta }(Q^2)}
e^{\Delta (x\to 0,Q^2)}  .$$ If
$$f(Q^2)\sim 1\ , $$
i.e. when $Q^2\gg Q_1^2$, we get the standard (Pomeron-dominated)
Regge behavior (with a $Q^2$ dependence in the effective Pomeron
intercept)
$$F_{2}^{(S,0)}(x\to 0,Q^2) \to A\ \left({Q^2\over Q^2 +
a}\right)^{1+\widetilde{\Delta}(Q^2)}\ \left({x_0\over x}
\right)^{\widetilde{\Delta}(Q^2)} \ \propto
x^{-\widetilde{\Delta}(Q^2)} .$$

The total cross-section for $(\gamma,p)$ scattering as a function
of the center of mass energy $W$ is
$$ \sigma ^{tot,(0)}_{\gamma,p} (W)=
4\pi^2\alpha\ \left[{F_{2}^{(S,0)}(x,Q^2)\over Q^2}\right]_{Q^2\to
0}  =\ 4\pi^2\alpha\ A\  a^{-1-\epsilon}\ x_0^\epsilon\
W^{2\epsilon}. $$

\begin{figure}[h]
\begin{center}
\hspace{-1.cm}
\includegraphics[width=0.64\textwidth]{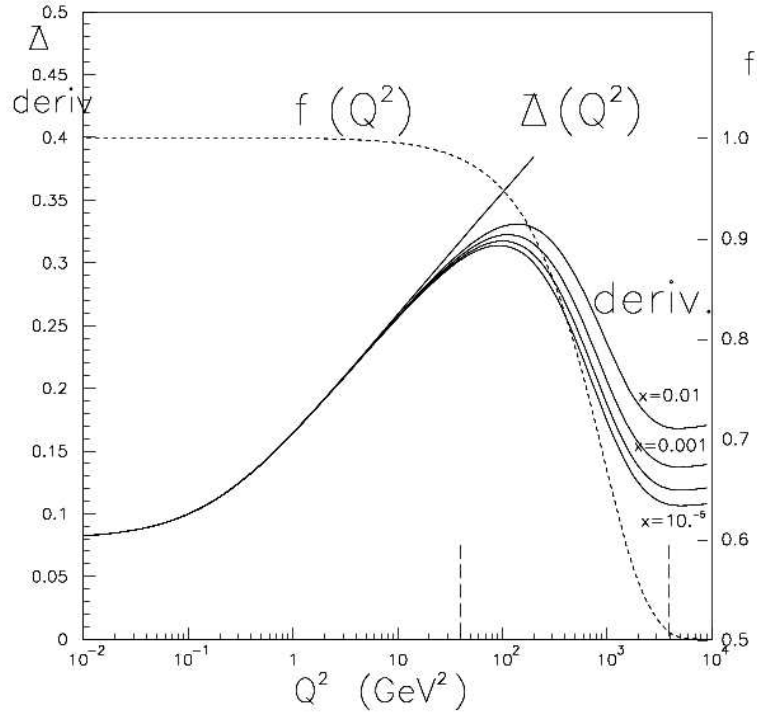}
\caption{\small \it {Logarithmic derivative of the proton SF,
$B_x(x,Q^2)=\frac{\partial \ln F_2(x,Q^2)}{\partial(\ln(1/x))}$
calculated in Ref. \cite{Interpolate} (see the text below).}}
\label{fig:Satur}
\end{center}
\end{figure}

Interestingly, in this model the logarithmic derivative of the
structure function,
$$B_Q(x,Q^2)={\partial F_2(x,Q^2)\over{\partial(\ln Q^2)}},\ \ \
B_x(x,Q^2)={\partial F_2(x,Q^2)\over{(\partial\ln (1/x))}},
$$
called $B_x$ slope, the equivalent of the Pomeron intercept in the
Regge domain, has a maximum, indicative of saturation, and shown
in Fig. \ref{fig:Satur}.


Since the saturation phenomena occur in the small-$x$, which is a
gluon-dominated region, we shall be mainly interested in that
component of the SF, namely in
$$F_2^S\sim xG(x,Q^2),$$
where $G$ is the gluon (singlet) distribution function, defined
above.

A slightly different, although related  approach is that of string
or parton percolation \cite{Pajares, Satz, JTT}. Gas-liquid phase
transition with a percolation type model was considered recently
in Ref. \cite{China}, however in that paper the "liquid" formed of
coloured quarks and gluons is colourless, i.e. hadronic.

\section{Statistical Models of SF}\label{Stat}
We assume that the interior of a nucleon (the idea can be extended
to nuclei as well), as seen in (inclusive) deep inelastic
scattering (DIS) or (exclusive) deeply virtual Compton scattering
(DVCS), is a thermodynamic system, that, similar to the case of
nuclear or heavy-ion collisions, bear collective (thermodynamic)
properties governed by a relevant equation of state (EoS). The
idea that DIS structure functions (SF) can be treated
thermodynamically by means of statistical mechanics is not new
\cite{Bhalerao, Cleymans, Soffer, JTT, Dubna}, moreover it
continues attract attention, although several subtle points remain
to be clarified. Above all, it concerns the choice of the relevant
coordinate system and of the corresponding variables.

For simplicity, we focus our attention on the small-$x$ singlet
(gluon) component of the SF, the extension to low-$x$ and/or the
non-singlet (valence quark) contributions being straightforward
\begin{equation}\label{xG}xG(x,Q^2)\sim {\frac{X_0x^b}{\exp{[(x-X_0)/\bar x]}+1}},\end{equation}
where $x$ is the Bjorken (light-cone) variable, $X_0$ is the
 chemical potential, that for the gluon component
can be set zero, and $\bar x$ is interpreted as the temperature
inside the proton.

In some papers, the dimensional energy $E$ (or momentum $k$)
variable is used in the statistical model of the SF instead of $x$
as in Eq. (\ref{xG}). This is not a simple kinematical problem,
since thermodynamics implies the presence of the dimensional
temperature in the statistical distribution like $k/T$ (be it of
the Fermi-Dirac, Bose-Einstein or Boltzmann type), while the
appearance of $x$ as in Eq. (\ref{xG}) needs some extra
modification. In Ref. \cite{Dubna} this was circumvented by using
a dimensionless "temperature" $\bar x=2T/m,$ where $m$ is the
proton mass, which is a consequence of the transition from the
rest frame to the infinite-momentum frame (IMF). Accordingly,
\begin{equation}\label{G}G(x)\sim\exp\Bigl{(-\frac{mx}{2T}}\Bigr).\end{equation}

The Boltzmann factor in the denominator of Eq. (\ref{xG}) can
mimic the large-$x\ \ (1-x)^n$ factor in the SF, although it
should be reconciled also with the quark counting rules, appearing
in the power $n$.

The last point is connected also with the EoS expected from the
statistical distribution of the type (\ref{xG}). Let us remind
that for an ideal gas of particles
$$P(T)=\int_0^\infty k d^3k \exp(-k/T),$$ which, due to radial
symmetry, can be rewritten as $$\int_0^\infty k^3 dk \exp(-k/T),$$
and by the change of variable $y=k/T$, one trivially arrives at
the Stafan-Boltzmann (S-B) EoS, $P\sim T^4$. This fact can be
interpreted also physically: the large-$x$ component of the SF
corresponds to a dilute perfect gas of partons. The low-$x$ factor
in Eq. (\ref{xG}) will affect the ideal Stefan-Boltzmann EoS only
when it will be written in Eq. (\ref{xG}) as $(x/\bar
x)^b=({mx\over{2T}})^b$ instead of $x^b.$ As a consequence, the
ideal Stefan-Boltzmann EoS will be modified as
\begin{equation}\label{P}P(T)\sim
T^{4+b}.\end{equation} The relative contribution of this
correction is negligible at small x, but it increases with $Q^2$
and decreasing $x$, resulting in a gas-liquid phase transition.

In most of the paper on the subject, $Q^2$ dependence is neglected
- either for simplicity, or "conceptually", by assuming that the
statistical approach applies to the SF for some fixed, "input"
value of $Q^2,$ from which it evolves according to the DGLAP
equation. We do not exclude high $Q^2$ evolution of the SF,
however with the following caveats:

a) the structure functions show strong $Q^2$ dependence, already
at low $x$, below the perturbative DGLAP domain; b) at large
$Q^2$, instead of the monotonic DGLAP evolution, due to the
proliferation of partons, the inverse process of their
recombination is manifest. This process is essential in our
interpretation of the saturation as a gas-liquid phase transition
(see the next Section). So, we prefer to keep explicit $Q^2$
dependence for all $x$ and $Q^2$. It is mild in the "gaseous"
region of point-like partons (at large $x$), becoming important
towards the saturation region (depending on both $x$ and $Q^2$),
where the point-like partons are replaced by finite-size droplets
of the partonic "fluid". This transition will be treated in the
next two sections by means of the classical van de Waals equation.

\section{Gas-Fluid phase transition in the van der Waals equation of
state}\label{IV}
 Having defined the statistical properties of the
SF, we now proceed to an equation of state (EoS) describing the
transition between a parton gas, via a mixed foggy phase, to the
partonic liquid. To this end, we use the van der Waals equation
\begin{equation}\label{VdW}(P+N^2a/V^2)(V-Nb)=NT,\end{equation} see, e.g. \cite{Fermi, Landau},
where $a$ and $b$ are parameters depending on the properties
of the system, $N$ is the number of particles and $V$ is the
volume of the "container", $V(s)=\pi R^3(s),\ \  R(s)\sim \ln s$
is the nucleon radius in our case. For point-like particles
(perfect gas), $a=b=0,$ and Eq. (\ref{VdW}) reduces to $pV=NT,$
and, since $N/V \sim T^3$, we get in this approximation $p\sim
T^4,$ to be compared with $p\sim T^{(4+b)},$ Eq. (\ref{P}), of the
preceding Section.

Alternatively, Eq. (\ref{VdW}) can be written as \cite{Fermi}
$$(P+a/V^2)(V-b)=RT, $$
or, equivalently
$$P=\frac{RT}{V-b}-\frac{a}{V^2}.$$
 The parameter $b$ is responsible for the finite dimensions of
the constituents, related to $1/Q$ in our case, and the term
$a/V^2$ is related to the (long-range) forces between the
constituents. From this cubic equation in $V$ one finds
\cite{Fermi} the following values for the critical values $V=V_c,\
\ P=P_c,\ \ $ and $T=T_c$ in terms of the parameters $a$ and $b$:
$$V_c=3b,\ \ p_c=a/(27b^2)\ \  T_c=8a/(27Rb).$$

The particle number $N(s)$ can be calculated as \cite{JS}
$$N(s)=\int_0^1dxF_2(x,Q^2),$$
where $F_2(x,Q^2)$ is given by Eqs. (1) or (\ref{Inter}). We
remind the kinematics: $s=Q^2(1-x)/x+m^2,$ which at small $x$
reduces to $s\approx Q^2/x.$ The radius of the constituent as seen
in DIS is $r_0\sim 1/Q,$ hence its two-dimensional volume is $\sim
Q^{-2}.$

By introducing "reduced" volume, pressure and temperature,
$${\cal P}=P/P_c,\ \ {\cal V}=V/V_c=\rho_c/\rho,\ \  {\cal T}=T/T_c, $$
the van der Waals equation (\ref{VdW}) can be rewritten as
\begin{equation}\label{VdWCr}\Bigl({\cal P}+3/{\cal V}^3\Bigr)\Bigl({\cal
V}-1/3\Bigr)=8{\cal T}/3.\end{equation}

Note that Eq. (\ref{VdWCr}) contains only numerical constants, and
therefore it is universal. States of various substances with the
same values of ${\cal P},\  {\cal V}$ and ${\cal T}$ are called
"corresponding states"; equation (\ref{VdWCr}) is called the "van der Waals
equation for corresponding states". The universality of the
liquid-gas phase transition and the corresponding principle are
typical if any system with short-range repulsive and long-range
attractive forces. This property is shared both by ordinary
liquids and by nuclear matter. This was demonstrated, in
particular, in Ref. \cite{Jaqaman}, where typical van der Waals
curves were shown to be similar to those derived by means of the
Skyrme effective interaction, see Fig. \ref{Jaqaman}.

Following Ref. \cite{Jaqaman}, we present two example of EoS, one
based on the Skyrme effective interaction and finite-temperature
Hartree-Fock theory, and the other one is the van der Waals EoS.
Jaqaman et al. \cite{Jaqaman} start with the EoS
\begin{equation}\label{Jaq1}
P=\rho kT-a_0\rho^2+a_3(1+\sigma)\rho^{(2+\sigma)},\end{equation}
where $\rho=N/V$ is the density and $a_0,\ \ a_3$ and $\sigma$ are
parameters, $\sigma=1$ corresponding to the usual Skyrme
interaction. Particular values of the above parameters,
corresponding to various options (degenerate and non-degenerate
Fermi gas) as well as to finite and infinite nuclear matter are
quoted in Table 1 of Ref. \cite{Jaqaman}.

According to the law of the corresponding states, Eq. (\ref{Jaq1})
is universal for scaled (reduced) variables, for which, with
$\sigma=1,$ it becomes
$$P=3{\cal  T}/{\cal V}-3/{\cal V}^2+1/{\cal V}^3,$$
to be compared with the van der Waals EoS
$$P=8{\cal  T}/(3{\cal V}-1)-3/{\cal V}^2.$$

\begin{figure}[h]
\begin{center}
\hspace{-1.cm}
\includegraphics[width=0.8\textwidth,angle=0]{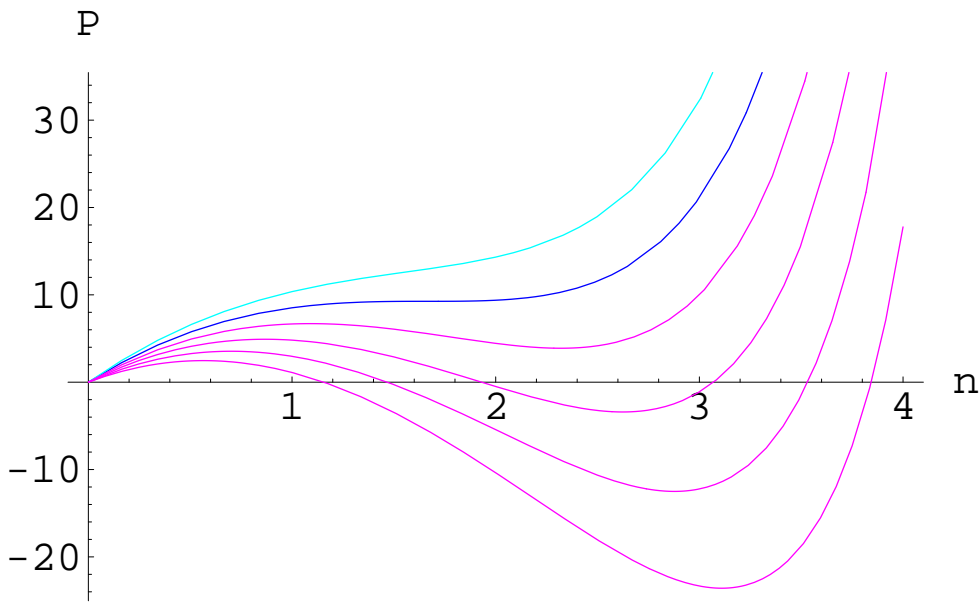}
\caption{\small \it {The pressure-to-density dependence
calculated, in arbitrary units, from Eq.(\ref{VdW11}).}}
\label{VdW1}
\end{center}
\end{figure}

Let us now write the van der Waals EoS in the form
\begin{equation}\label{VdW11}
P(T;N,V)=-\Bigl(\frac{\partial F}{\partial
V}\Bigr)_{TN}=\frac{NT}{V-bN}-a\Bigl(\frac{N}{V}\Bigr)^2=
\frac{nT}{1-bn}-an^2,
\end{equation}
where $n=N/V$ is the particle number density, $a$ is the strength
of the mean-field attraction, and $b$ governs the short-range
repulsion. We identify the particle number density with the SF
$F_2(x,Q^2)$ of Sec. \ref{II}. Fig. \ref{VdW1} shows the
pressure-density dependence calculated from Eq. (\ref{VdW11}) with
$a=5\ GeV^{-2}$ and $b=0.2\ GeV^{-3}$. Notice that while Eq.
(\ref{VdW11}) is sensitive to $b$ (short-range repulsion), it is
less so to $a$ (long-range attraction). Representative isotherms
are shown in this figure: the dark blue line (second from the top)
is the critical one, $T_c=8a/(27b)$. Above this temperature (top
line, in pale blue), the pressure rises uniformly with density,
corresponding to a single thermodynamical state for each $P$ and
$T$. By contrast, for subcritical temperatures ($0<T<T_c$) (red
lines) the function $P(n)$ has a maximum followed by a minimum,
see Fig. \ref{VdW1}. Below the critical value $P_c,$ three density
regimes exist \cite{Randrup}. The smallest density region lies in
the the gaseous phase below the spinodal region, while the highest
densities lie in the liquid phase, above the spinodal region. The
coexistence phase can be determined by a Maxwell construction.

In the next section we attempt to match the parameters appearing in
the van der Waals equation with those of the nucleon structure
functions. To do so, we use also an equation of state derived from
the $S$-matrix formulation of statistical physics which, similar
to the van der Waals EoS, contains metastable states.

\section{Mapping the saturation region in DIS onto the spinodal region in the VdW EoS;
metastability, overheating and supercooling}\label{Mapping}

The correspondence between the EoS with its  variables $P,\ T,\
\mu$ {\it etc} and the observables, depending on the reaction
kinematics, is the most delicate and complicated point in the
thermodynamical description of any high-energy collisions,
especially of DIS. It needs caution, further studies and numerical
tests. Attempts to link two different approaches to hadron
dynamics, one based on the $S$ matrix (scattering amplitude, cross
sections) and the other one on their collective properties
(statistical mechanics, thermodynamics, equation of state) are
known from the liturature \cite{DMB,Shurik}.

\begin{figure}[h]
\begin{center}
\includegraphics[width=\textwidth,angle=0]{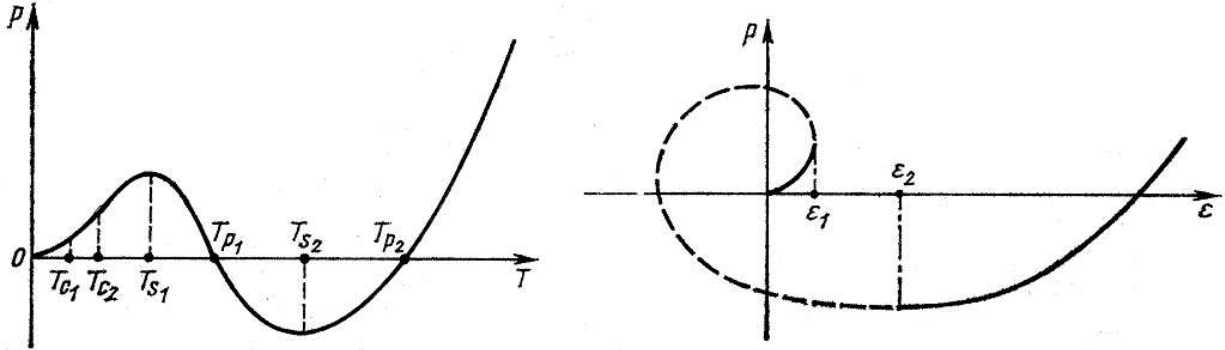}
\caption{\small \it {$P(T)$ and $P(c)$, where $c$ is the sound
velocity, dependence calculated from Eq.(\ref{Shurik}).}}
\label{6}
\end{center}
\end{figure}

\begin{figure}[h]
\begin{center}
\includegraphics[width=\textwidth]{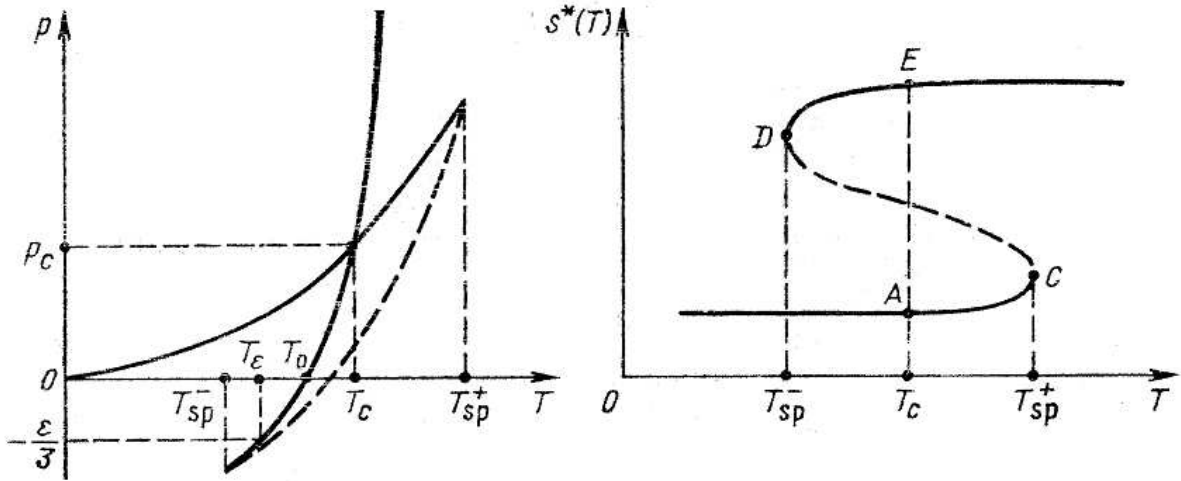}
\caption{\small \it {Metastable super-cooled and over-heated
states calculated from a modification of the Bag EoS, see
\cite{EChAYa}).}} \label{7}
\end{center}
\end{figure}

An original EoS of the ultra-relativistic nuclear matter matter
based on the $S$ matrix formulation of statistical physics
\cite{DMB} was derived in Ref. \cite{Shurik}. For vanishing
chemical potential, $\mu=0$, it reads
\begin{equation}\label{Shurik}
    P(t)=AT^4-BT^5+ T^6,
\end{equation}
where the values of the coefficients $A,\ B$ and $C$ (all
positive!) were determined \cite{Shurik} by the hadron scattering
data. Asymptotically, $T>>m,\
 P(T)\sim T^6,$ where $m$ is the proton mass. It differs from the
 expected (but not confirmed) properties of the so-called quark-gluon
 plasma predicted by perturbative QCD. The $P(T)$ dependence of the EoS (\ref{Shurik})
 is shown in Fig. \ref{6}. Similar to the van der
 Waals curve, Fig. \ref{Jaqaman} or Fig. \ref{VdW11}, it has a maximum $T_{s_1}$ followed by a minimum ($T_{s_2}$).
 At the minimum, the pressure is negative and the system is
 metastable. Before reaching the minimum at $T_{s_2}$ (from the left), the system
 will undergo a phase transition to $T_{c_2}$ of the "cooler"
 branch. The interval between these points is unphysical.
 Metastable super-cooling and over-heating within the bag EoS is
 shown in Fig. \ref{7} from \cite{EChAYa}. Similarities
 between the VdW EoS and Eq. (\ref{Shurik}) were quantified in Ref.
 \cite{Shurik1}, where the parameters $a$ and $b$ appearing in the van der Waasl EoS
 were assumed to be temperature-dependent, $a=a_0/T^\alpha,\ \
 b=T^\beta.$ Note that the limiting particle density is $1/b$.

 Eq. (\ref{Shurik}) was derived \cite{Shurik} from an on-shall
 hadronic scattering amplitude ($Q^2=m^2$). Its off-shell modification, based
 on DIS or DVCS, might tell us much more about the connection
 between structure functions (partonic distributions) and the corresponding EoS.

\section{Discussion and conclusions} \label{Concl}
Below we list our (temporary) conclucions and mention some open
questions left behind this paper:

$\bullet$ The aim of our paper is not just another parametrization
of DIS structure functions; instead we propose a new insight into
the properties of the interior of the nucleon.

$\bullet$ For simplicity, we have concentrated on the singlet
component of the SF (gluons and sea quarks), dominating the
low-$x$ region, where saturation takes place. The inclusion the
non-singlet and higher-$x$ components (valence quarks), according
to the prescriptions given in Sec. \ref{II}, are straightforward.
Relevant fits to the data can be found in Refs. \cite{Bhalerao,
Cleymans, Soffer, Dubna} and will be treated elsewhere.

$\bullet$ We omitted any discussion of the large-$x\ \ (1-x)^n$
factor in the SF. This is because the statistical approach is not
valid in the $x\rightarrow 1$ limit, on the one hand, and for
simplicity, on the other hand.

$\bullet$ We did not consider the possible (a)symmetry between
"heating" and "cooling" of the excited nucleon. During deep
inelastic scattering, the nucleon (nucleus) gets excited (heated),
whereafter it releases its energy (heat) by producing secondaries.
The energy distribution of these secondaries may reveal the
temperature of the system, however this "ultimate" temperature
\cite{Hagedorn}, created after the confinement transition, will be
different from that of the partonic system inside the proton,
which may increase indefinitely.

$\bullet$ Any phase transition may produce fluctuations in the
observed spectra of produced particles. These fluctuations may
originate either from the gas-fluid-gas phase transition under
discussion, or from the (de)confinement transition, beyond the
scope of the present paper.

$\bullet$ The gas-fluid transition does not necessarily follow the
van der Waals EoS. Possible alternatives are e.g. percolation
\cite{Pajares, Satz} or clustering of partons similar to the case
of the molecule.
 A possible, molecule-like  mechanism for
hadron aggregation, was proposed for crossover compatible with
color confinement  in Ref. \cite{China}, and a toy model to
realize this mechanism was constructed. A kind of grape-shaped
quark gluon plasma (gQGP) was obtained. Pair distribution
functions for gQGP were calculated exhibits the character of a
liquid. Quark recombination or coalescence, extending the concept
of single parton fragmentation, which has been used in elementary
collisions since the '70s, was studied recently in Ref.
\cite{Becattini} in the context of heavy ion collisions. It should
be remembered however, that the above models include
(de)confinement transition, absent from our approach.

$\bullet$ Experimentally, the onset of the gas-fluid phase
transition may be verified by the observed spectrum of the
produced particles. The $p_\perp$ distribution of produced from
the dilute gaseous state, i.e. below the saturation region, can be
computed perturbatively, while beyond the saturation border line
they result from the collision of a very large number of
constituent, as in the color glass condensate \cite{glass}. The
observation of any correlation between the transverse distribution
of the particles produced in DIS (or DVCS) below- or beyond the
saturation border line (see Sec. \ref{II}) will bring evidence for
or against the picture presented in this paper.

$\bullet$ Whatever the details of the transition, the important
point to realize is the existence of a dense partonic substance,
different from the perfect partonic gas, associated with Bjorken
scaling in DIS or the so-called quark-gluon plasma, predicted by
perturbative QCD and expected in high-energy hadronic and/or
nuclear collisions. Instead, experimental data on DIS and on
high-energy heavy ion collisions show that the partonic matter may
appear as a fluid, called  by L. McLerran {\it et al.} "color
glass condensate" \cite{glass}. The nature of the strongly
interacting matter under extreme conditions should be universal,
be it produced in hadron-hadron, heavy nuclei or in deep inelastic
lepton-hadron scattering.

$\bullet$ A general remark, concerning collective properties of
the nuclear matter: matter is made of quarks, while gluons are
binding forces between them. In that sense, strictly speaking,
statistics should be applied to quarks rather than gluons.

\section*{Acknowledgments}

We thank F.~Becattini, M.~Gorenstein,  K.~Kutak, V.~Magas,
F.~Paccanoni and A.~Pimikov for discussions. A.~N. and J.~T are
grateful to A.~Lengyel, J.~Kontros, A.~Spenik and Z.~Tarics for
their assistance and support. We also thank Yura Stelmakh for his
help in preparing the manuscript. The work of L.~J. was supported
by the grant "Matter under Extreme Conditions" of the Department
of Astronomy and Physics of the National Academy of Sciences of
Ukraine.


\newpage
\begin{thebibliography}{99}

\bibitem{fermi} E.Fermi, {\it High Energy Nuclear Events}, Progr. Theor. Phys., {\bf 5} (1950)
70.

\bibitem{Bhalerao} R.S. Bhalerao, {\it Statistical model for the nucleon structure functions},
Phys. Lett. B {\bf 380} 1 (1996), hep-ph/9607315; R.S. Bhalerao
and R.K. Bhaduri, {\it Droplet formation of quark-gluon plasma at
low temperatures and high high densities}, hep-ph/0009333; R.S.
Bhalerao, N.G. Kelkar and B.~Ram {\it Model for polarized and
unpolarized parton density functions in the nucleon}, Phys. Rev. C
{\bf 63} 025208 (2001), hep-ph/9911286; K. Ganesamurtly, V.~
Devanathan and M. Rajasekaran, Z. Phys. C {\bf 52} 589 (1991; V.
Devanathan, S. Karthiyaini and K. Ganesamurthly, Mod. Phys. Lett.
A {\bf 9} 3455 (1994); V. Devanathana and J.S. McCarthy, Mod.
Phys. Lett. A {\bf 11} 147 (1996); Hai Lin, hep-ph/0105050,
hep-ph/0105172, hep-ph/0106100.

\bibitem{Cleymans} C. Angelini and R. Pazzi, Phys. Lett. B {\bf 135} 473 (1984).
E. Mac and E. Ugaz, Z. Phys. C {\bf 43} 655 (1989); J. Cleymans
and R.L. Thews, Z. Phys. C {\bf 37} 315 (1988); J. Cleymans, I.~
Dadic, and J. Joubert, Z. Phys. {\bf 68} 275 (1994).

\bibitem{Soffer} C. Bourrely, F. Buccella, G. Miele, G. Migliore,
J. Soffer, and V. Tibullo, Z.Phys. C {\bf 62} 431 (1994); C.
Bourrely and J. Soffer, Phys. Rev. D {\bf 51} 2108 (1995); C.
Bourrely and J.~Soffer, Nucl. Phys. B {\bf 445} 341 (1995); Claude
Bourrely, Franco Buccella, Jacque Soffer, hep-ph/1008.5322.

\bibitem{JTT} L. Jenkovszky, S. Troshin and N. Tyurin, In the
Proc. of the EDS ("Blois") Canference, held at CERN, 2009;
E-print, 2009, hep-ph/1002.3527; L. Bulavin, L. Jenkovszky,
 S. Troshin and  N. Tyurin, Physics of Particles and Nuclei,
 {\bf 41} 924 (2010), Contribution to the Bogolyubov 2009
Conference, held in Dubna, August, 2009, ; L. Jenkovszky, S.
Troshin, and N. Tyurin, {\it Critical phenomena in DIS}, In: {\it
Gauge Fields. Yesturday. Today. Tomorrow. Slavnov Fest}, Moscow
Steklov Institute Proceedings, 2010, in press.

\bibitem{Dubna} J. Cleymans et al. E-print, 2010. hep-ph/1004.2770.

\bibitem{Hagedorn} R.Hagedorn, {\it Remarks on thethermodynamical model of strong interactions},
Nucl. Phys., {\bf B24} (1970), 93.

\bibitem{Krzywicki} Andr\'e Krzywicki, LPT-ORSAY-02-31 preperint, Apr. 2002, E-print hep-ph/020411.

\bibitem{Yoshida} R. Yoshida, {\it What HERA can tell us about
saturation}, In the proceedings of the ISMD08 Conference, DESY,
2008.


\bibitem{Jaqaman} H. Jaqaman, A. Z. Mekjian, and L. Zamik {\it Nuclear condensation}, Phys.
Rev. C  {\bf 27} (1983) 2782-2791.


\bibitem{Rivista} M. Bertini, M. Giffon, L.L. Jenkovszky,
Paccanoni F., and Predazzi E., Rivista Nuovo Cim.  {bf 19} (1996)
1-45.

\bibitem{TT}  S. Troshin and N. Tyurin, Mod. Phys. Lett. A {\bf  23} (2008) 3141;
E-print, 2008. hep-ph/0803.1917.


\bibitem{Jenk}  L. Jenkovszky {\it Dual analytic model of generalized parton
distributions}, Yad. fiz. {\bf 71} (2008) 372-384.

\bibitem{Golec}  A.M. Sta\'{s}to,  K. Golec-Biernat and  J. Kwieci\'{n}ski, Phys. Rev. Letters, {\bf 86} (2001)
596; Krzysztof Golec-Biernat, E-print, 2009. hep-ph/0109010.

\bibitem{BFKL} E.A. Kuraev, L.N. Lipatov, V.S. Fadin, Sov. Phys. JETP {\bf 45} (1977) 199;
I.I. Balitsky, L.N.~Lipatov, Sov. J. Nucl. Phys. {\bf 28} (1978)
822.

\bibitem{Kovcheg} Yuri V. Kovchegov, {\it Introduction to the
Physics of Saturation}, arXiv:1007.5021.

\bibitem{BK} I. Balitsky, Nucl. Phys. {\bf B463} (1996) 99; arXiv:
hep-ph/9901281; Y.V. Kovchegov, Phys. Rev. {\bf D60} (1999)
034008; arXiv:hep-ph/9901281.

\bibitem{Gelis}  F. Gelis, J. Phys. {\bf 34} (2007) S421; hep-ph/0701225.


\bibitem{Interpolate} P. Desgrolard, L. Jenkovszky, and
 F.~Paccanoni, EPJ  {\bf C7} (1999) 263.




\bibitem{Pajares} C. Pajares, E-print, hep-ph/050111125.


\bibitem{Satz} P. Castorina, K. Redlich and H.~Satz, {\it The
Phase Diagram of Hadronic Matter}, BI-TP 2008/17 preprint,
hep-ph/0807.4469.


\bibitem{China} Xu Mingmei, Yu Meiling and Liu Lianshou,
{\it Mechanism of crossover between hadron gas and QGP and the
liquid property of sQGP}, Nuclear Physics {\bf A 820} (2009)
131–134.

\bibitem{Fermi}  E. Fermi, {\it Termodinamica}, Boringhieri, Torino,
1958;


\bibitem{Landau} L.D.  Landau and E.M. Lifshitz, {\it Statisticheskaya
Fizika}, Part 1, Nauka, Moscow, 1976.

\bibitem{JS} L.L. Jenkovszky, B.V. Struminsky, ITP preprint,
Kiev,1977.

\bibitem{Randrup} Philipp Chomaz, Marin Colonna, J{\rm$\o$}rgen Randrup,
Phys. Rep. {\bf 389} (2004) 265.

\bibitem{DMB} R. Dashen, S. Ma, H.J. Bernstein, Phys. Rev. {\bf
187} (1969) 345.

\bibitem{Shurik} L.L. Jenkovszky, A.N. Trushevsky, Nuovo Cim. A {\bf 34} (1976) 360;
L.L. Jenkovszky, A.N.~Shelkovenko, ibid, {\bf 101} (1989) 137.

\bibitem{EChAYa} V.G. Bojko, L.L. Jenkovszky, V.M. Sysoev, EChAYa
(PEPAN), {\bf 22} (1991) 675.


\bibitem{Shurik1} A.I. Bugrij, A.N. Trushevsky, {\it The van der Waals equaton of state
for the system of ultrarelativistic particles}, Preprint
ITP-78-82E,Kiev, 1978.

\bibitem{Becattini} Francesco Becattini and R. Fries, {\it The QCD
confinement transition: hadron formation}, nucl-th/0907.1031.

\bibitem{glass} For a recent review see: Fran\c{c}ois Gelis, {\it
Color Glass Condensate and Glasma}, hep-ph/1009.0093.


\end{thebibliography}
\end{document}